\documentclass[conference]{IEEEtran}
\IEEEoverridecommandlockouts

\usepackage{cite}
\usepackage{amsmath,amssymb,amsfonts}
\usepackage{algorithmic}
\usepackage{graphicx}
\usepackage{algorithm}
\usepackage{textcomp}
\usepackage{xcolor}
\usepackage{subfig}
\usepackage{multirow}
\usepackage{todonotes}
\def\BibTeX{{\rm B\kern-.05em{\sc i\kern-.025em b}\kern-.08em
    T\kern-.1667em\lower.7ex\hbox{E}\kern-.125emX}}
\begin{document}

\title{RRC Signaling Storm Detection in O-RAN}

\author{\IEEEauthorblockN{Dang Kien Nguyen}
\IEEEauthorblockA{\textit{Standards \& Technology} \\
\textit{Ericsson France}\\
dang.kien.nguyen@ericsson.com}
\and
\IEEEauthorblockN{Rim El Malki}
\IEEEauthorblockA{\textit{Standards \& Technology} \\
\textit{Ericsson France}\\
rim.el.malki@ericsson.com}
\and
\IEEEauthorblockN{Filippo Rebecchi}
\IEEEauthorblockA{\textit{Standards \& Technology} \\
\textit{Ericsson France}\\
filippo.rebecchi@ericsson.com}
}

\maketitle

\begin{abstract}
The Open Radio Access Network (O-RAN) marks a significant shift in the mobile network industry. By transforming a traditionally vertically integrated architecture into an open, data-driven one, O-RAN promises to enhance operational flexibility and drive innovation. In this paper, we harness O-RAN’s openness to address one critical threat to 5G availability: signaling storms caused by abuse of the Radio Resource Control (RRC) protocol. Such attacks occur when a flood of RRC messages from one or multiple User Equipments (UEs) deplete resources at a 5G base station (gNB), leading to service degradation. We provide a reference implementation of an RRC signaling storm attack, using the OpenAirInterface (OAI) platform to evaluate its impact on a gNB. We supplement the experimental results with a theoretical model to extend the findings for different load conditions. To mitigate RRC signaling storms, we develop a threshold-based detection technique that relies on RRC layer features to distinguish between malicious activity and legitimate high network load conditions.
Leveraging O-RAN capabilities, our detection method is deployed as an external Application (xApp). Performance evaluation shows attacks can be detected within 90ms, providing a mitigation window of 60ms before gNB unavailability, with an overhead of 1.2\% and 0\% CPU and memory consumption, respectively.

\end{abstract}


\section{Introduction}

O-RAN (Open Radio Access Network) is a 5G architecture that emphasizes openness, flexibility, and interoperability by disaggregating traditional RAN components and using standardized interfaces~\cite{polese2023understanding}. One of its key capabilities is allowing applications to be plugged-in for various functions, including the detection and mitigation of security threats, offering a modular approach to optimizing network performance~\cite{niknam2022intelligent}. Leveraging cloud-native principles and AI/ML (Artificial Intelligence/Machine Learning) technologies, O-RAN enhances dynamic resource allocation and real-time network adaptation, making it a vital part of the next-generation telecommunication ecosystem~\cite{upadhyaya2023open,damnjanovic2024spectrum}.

In 5G networks, RRC signaling storms refer to a form of Denial-of-Service (DoS) attack (i.e., against the control plane) where a malicious User Equipment (UE) overwhelms the network by sending a surge of RRC connection requests, depleting available resources and potentially causing service degradation or outages (i.e., unavailability of the base station, hence preventing new UEs from reaching the RRC connected state). These storms exploit vulnerabilities in the signaling process, particularly in the RRC layer, which handles connections between the UE and gNodeB (gNB)~\cite{tabiban2023signaling}.


Existing detection solutions focus on higher-layer features and do not adequately consider low-layer characteristics of the RRC protocol. Furthermore, they overlook specific causes, such as emergency or high-priority establishment requests, and fail to differentiate between genuine high-load scenarios and malicious attacks (i.e., since both may have similar traffic patterns), making them less effective in real-world conditions where such distinctions are critical~\cite{hoffmann2023signaling,feng2023research}.

\noindent
In this paper, we have two main objectives:
\begin{enumerate}
    \item to implement and test the impact of RRC signaling storms using OpenAirInterface (OAI);
    \item to propose and evaluate a detection technique for these attacks in 5G networks based on the O-RAN architecture.
\end{enumerate} 
In particular, we leverage on O-RAN's capability to integrate applications (e.g., rApps, xApps) for detection. Our proposed solution requires to differentiate accurately between attack scenarios and legitimate high-load conditions using RRC layer features. We test the detection method in the most challenging cases, e.g., when UEs use a random identity to register and with emergency or high-priority establishment causes. In such cases, it is challenging to determine whether requests originate from a single Malicious UE (MUE) or multiple Benign UEs (BUEs). Additionally, the prioritization of these establishment causes allows a MUE to reserve RRC resources more effectively, complicating detection.
To validate the experimental findings of the RRC signaling storm, we also propose a theoretical model that aligns with the observed attack behavior. The detection solution is evaluated for security and performance, proving its ability to distinguish between an attack and a high-load scenario before the gNB becomes unavailable (i.e., with an acceptable delay of $\approx 90$ms), with minimal resource overhead (i.e., $\approx 1.2\%$ CPU and $\approx 0\%$ memory consumption).

\section{Literature Review}\label{related}

The work in \cite{hoffmann2023signaling} introduces an O-RAN xApp for detecting abnormal activities in Industrial Internet of Things (IIoT) devices during registration by using O-RAN interfaces to gather long-term network statistics. However, this approach relies on statistical threshold-based detection, which limits its ability to differentiate between attacks and high-load scenarios. In \cite{feng2023research}, the authors combine time series prediction, adaptive thresholds, and anomaly detection to forecast signaling storms. While effective, this method lacks the inclusion of lower-layer features such as RRC, reducing its precision in detecting attacks targeting lower network layers. 
Similarly, \cite{ettiane2021toward} addresses 5G-RAN security challenges, focusing on two specific DoS attacks targeting RRC state transitions and fake system information requests. The proposed countermeasures include introducing randomness in system parameters to mitigate these threats. The randomization technique in \cite{ettiane2021mitigating} effectively reduces signaling volume in 5G RRC, demonstrating its efficacy in mitigating DoS signaling attacks. 
In \cite{escudero2019detecting}, the authors apply Dempster-Shafer theory to classify network operations as either attacks or normal activity. However, this method struggles with distinguishing high-load scenarios and is not applicable in cases involving random UE IDs. Meanwhile, the approach in \cite{abdelrahman2016detecting} employs a supervised RNN-based (Recurrent Neural Network) model to detect excessive RRC signaling by monitoring packets at the network edge, though it does not analyze lower radio layers. 
The research in \cite{pavloski2019detecting} uses thresholds on consecutive UE requests to identify low-volume attacks, but faces challenges with false positives and limited knowledge of UE identities. Lastly, \cite{xavier2023machine} utilizes OpenRAN combined with machine learning for early attack detection in the RAN, focusing on a set of features, other than those at the RRC layer, which remain underexplored. 

The existing literature insufficiently analyzes RRC layer features for detecting signaling storms, often lacks testing under high-load conditions, and provides vague threat models in complex scenarios like random UE IDs and emergency access. Section~\ref{sol} addresses these gaps with a threshold-based detection solution.

\section{Background}\label{background}

This section provides the necessary background information on O-RAN architecture, the 5G connection establishment process, and the OAI platform which has been used for attack implementation and detection evaluation.


\subsection{O-RAN Architecture}\label{bkoran}

The O-RAN architecture consists of~\cite{sivaraj2023ran}: 
\begin{itemize}
\item \textbf{Disaggregated components:} The architecture separates traditional base station functions into components such as the Central Unit (O-CU), Distributed Unit (O-DU), and Radio Unit (O-RU), which can be deployed flexibly. 
\item \textbf{RAN Intelligent Controller (RIC):} It is comprised of a Near-Real-Time RIC (for decisions within milliseconds) and a Non-Real-Time RIC (for long-term tasks).
\item \textbf{Service Management and Orchestration (SMO):} It ensures efficient resource management across O-RAN components.
\item \textbf{Open Interfaces:} Open, standardized interfaces for communication between network elements (e.g., O-RU, O-DU, and O-CU), ensuring interoperability between different vendors' equipment. 
\end{itemize}




\subsection{5G Connection Establishment}



\begin{figure}[!t]
\centering
\includegraphics[width=0.5\textwidth]{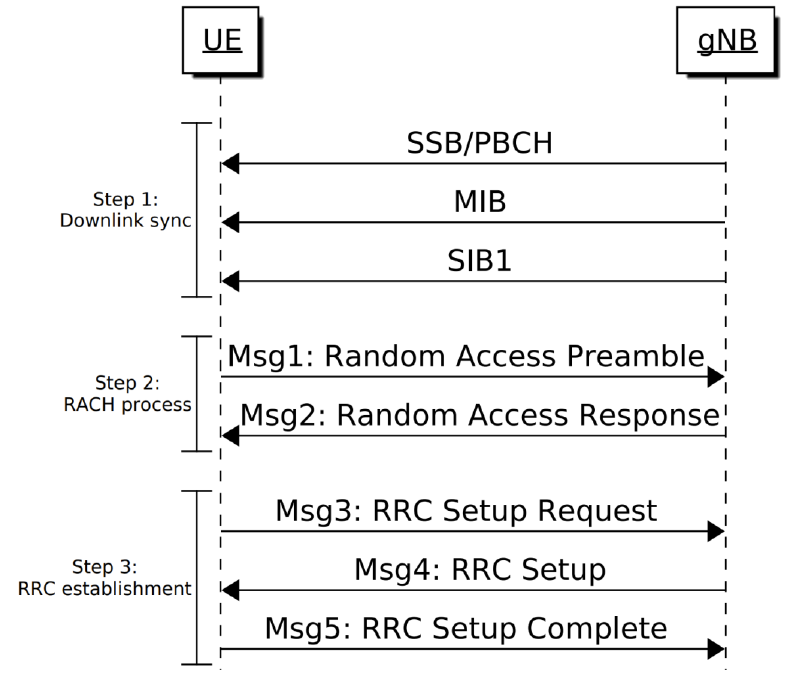}
\caption{5G connection establishment}
\label{est}
\end{figure}

The 5G connection establishment process allows UEs to connect to the network for services such as data and voice communication. This involves authentication, authorization, and configuration of the UE. Figure~\ref{est} details its key steps, e.g., the downlink synchronization (step 1), the RACH (Random Access Channel) procedure (step 2), and the RRC connection establishment procedure (step 3). The RRC procedure, managed by the CU at Layer 3, controls the connection between the UE and the gNB, overseeing the establishment, maintenance, and release of the radio link for efficient communication~\cite{hailu2018rrc,mukherjee20195g}.

\begin{itemize}
\item \textbf{RRC Setup Request:} 
The UE initiates the RRC procedure by sending an RRC setup request (Msg3) to the network after the RACH process. It includes the UE's identity, that can be S-TMSI (SAE Temporary Mobile Subscriber Identity) or a random value, and an establishment cause.

\item \textbf{RRC Setup:}
The gNB responds with an RRC setup message (Msg4), allocating resources and providing necessary configuration parameters. This includes assigning a radio bearer and specifying the initial settings for communication.

\item \textbf{RRC Setup Complete:} 
The UE confirms the setup by sending an RRC setup complete message (Msg5). This step finalizes the connection, allowing the UE to begin data transmission and other communication tasks (i.e., UE will be in the RRC\_CONNECTED state).
\end{itemize}
 
\subsection{OAI Platform}

OpenAirInterface (OAI) is an open-source software framework designed to provide a flexible and configurable implementation of 4G and 5G mobile communication systems~\cite{kaltenberger2020openairinterface}. OAI is modular and consists of the following elements: 
\textbf{OAI-UE}, 
\textbf{OAI-gNB} (i.e., a software-based implementation of a 5G base station gNB), 
\textbf{FlexRIC} (i.e., an open-source Near-RT-RIC which, in the proposed setup, extracts lower layer features such as RRC features from the OAI-gNB and relays them to the xApp)~\cite{schmidt2021flexric}, 
\textbf{xApp} (i.e., it can be used to perform different functions, e.g., the proposed detection solution), and OAI-5G-Core. The OAI-UE and OAI-gNB communicate over an RF-simulated (Radio Frequency) air interface. 

\section{RRC Signaling Storms}\label{attack}

In this section, RRC signaling storms are first defined, followed by a description of the considered test scenarios and the presentation of the theoretical model. 

\subsection{RRC Signaling Storm Definition}

RRC signaling storms impact the control plane's availability, leading to the inability for new UEs trying to connect to transition to the ``RRC\_CONNECTED state''~\cite{tabiban2023signaling}. In a nutshell, a signaling storm may occur when the rate of connection requests (i.e., Msg3s) that are received at the gNB exceeds a given capacity. Detecting and mitigating such malicious activities during the initial attachment procedure is challenging because they occur prior to authentication, and limited information is available at the gNB about the communicating UEs. For instance, a malicious UE (the attacker) can repeatedly cycle through Msg1 to Msg3, without waiting for the RRC procedure to complete (e.g., it will never respond back with a Msg5). If this loop is executed rapidly enough, it can deplete resources at the gNB, leading to its unavailability~\cite{zhang2024mitigating}.
An important factor to consider in the RRC procedure is the waiting time at the gNB, as it plays a crucial role in RRC signaling storms.




\textbf{Waiting time:} 
Upon receiving Msg3, the gNB typically allocates a context (resource) for the UE, responds with Msg4, and starts a timer to await Msg5. This timer, known as the ``waiting time'', can range from $2$ to $3$ seconds. In our setup, it is configured to $2.7$ seconds.
This means that the gNB (roughly) reserves UE context for the duration of the waiting time. In normal operation, the processing time between the reception of a Msg3 and the transmission of Msg4 is in the order of milliseconds. Once the waiting time expires, the gNB releases the reserved resources. Since the gNB has a limited amount of resources to be reserved for context, if all resources are occupied, the RRC processing at the gNB becomes unavailable. In other words, the attacker’s goal is to exhaust these resources by reserving them without ever completing the RRC process.

We also define the \textbf{drop time} as the moment when all available resources are consumed, rendering the gNB blocked or unavailable. Prior to the drop time, there is an availability period that we define as the \textbf{duration of accept}, during which the gNB can respond to UE connection attempts. Following the drop time is the \textbf{duration of reject}, defined as the time duration when the gNB has no available RRC resources and rejects incoming RRC messages. During this period, new UEs cannot connect, and the gNB is overloaded. Resources may become available again when at least one waiting time expires. However, in the event of an attack, these resources are quickly reoccupied after their release.


\subsection{Possible Scenarios}

We consider two main scenarios: 1) a deliberate attack, and 2) a high-load scenario. 

\noindent
\textbf{Attack scenario:} 
The attacker aims to exhaust all RRC resources at the gNB by sending Msg3s at a high rate, leading to a decrease in the gNB's availability.
Consequently, a drop in the availability of the gNB occurs.



By increasing the attack rate (i.e., more Msg3s per second), we expect the duration of accept to decrease, meaning the drop time occurs faster. Consequently, the duration of reject increases, amplifying the attack's impact. The severity of an attack is also directly related to the number of available RRC resources. As the number of available resources decreases — e.g., when more benign UEs are connected — the overload state is reached faster and maintained for prolonged periods, requiring less effort from the attacker. 

\textit{To induce an overload state, all available RRC resources must be occupied/reserved in less time than the waiting time (i.e., drop time $\textless$ waiting time), ensuring that the duration of reject is sustained. } 

\noindent
\textbf{High-load scenario:} It represents a legitimate surge in network activity due to an increased number of UEs or elevated traffic demand (e.g., flash mob event, natural disaster site, etc.). 

Available resources are also reserved faster than their waiting time, causing the gNB to enter an overload state. The key difference with an attack scenario is that benign UEs attempt to complete the RRC setup procedure, and several Msg5s are received by the gNB. Resources are only released when at least one benign UE disconnects.

It is important to differentiate between a legitimate high-load scenario and a malicious attack to ensure that appropriate responses are applied, maintaining network performance and security without unnecessarily disrupting service or misallocating resources.

\subsection{Theoretical Model}\label{theosec}


In addition to the practical implementation of the attack, we have developed a theoretical model of the behavior of a 5G gNB under an RRC signaling storm attack. This model aids in understanding the performance of a gNB's response to varying attack intensities and in predicting its behavior under different load conditions.

Based on the proposed model, six output metrics can be calculated/derived: number of accepted Msg3s ($N_{A}$), number of rejected Msg3s ($N_R$), drop time ($T_D$), duration of accept ($T_A$), duration of reject ($T_R$), and the overall gNB's availability rate ($R_{avai}$). To obtain these outputs, five input parameters are required: the waiting time ($T_W$), the max number of UEs supported/handled by the gNB simultaneously (gNB's capacity) ($N_{UE}$), the attacker's rate ($R_{att}$), the rate of benign UEs ($R_{BUE}$), and the number of connected (benign) UEs in the network ($N_{BUE}$).

The drop time, a critical moment when the gNB can no longer respond to new incoming Msg3s and becomes unavailable or blocked, can be calculated by dividing the current gNB capacity (i.e., the maximum capacity minus the number of connected UEs) by the attacker's rate:
\begin{align}
T_D&=\frac{N_{UE}-N_{BUE}}{R_{att}}, \quad &N_{BUE}\neq 0\\
T_D&=\frac{N_{UE}}{R_{att}}, \quad &N_{BUE} = 0
\end{align}

The duration of accept is equal to the drop time ($T_A=T_D$), hence, the duration of reject can be derived as follows: $T_R=T_W-T_A=T_W-T_D$.

To calculate the number of accepted Msg3s, the number of connected (benign) UEs is subtracted from the gNB's maximum UE capacity:
\begin{align}
N_A&=N_{UE}-N_{BUE}
\end{align}

If no BUEs are connected to the gNB, then $N_A=N_{UE}$.



Having calculated the duration of reject, the number of rejected Msg3s can be determined by multiplying the duration of reject by the rate of incoming Msg3s (considering the rates of the attacker and BUEs): 
\begin{align}
N_R&= T_R * (R_{att} + R_{BUE})
\end{align}
In the case of an attack, the attacker's rate is much higher than the rate of BUEs ($R_{att} \gg R_{BUE}$). Thus, the rate of BUEs becomes negligible. Consequently, the number of rejected Msg3s can be approximated as follows:
\begin{align}
N_R \approx T_R * R_{att}
\end{align}

Finally, the gNB's availability ($R_{avai}$) and unavailability ($\overline{R_{avai}}$) rates can be obtained as follows:
\begin{align}
&R_{avai}= \frac{\sum^{N_{rep}}_{i=1} N_A^i}{\sum^{N_{rep}}_{i=1} {N_A^i+N_R^i}}*100,\\
&\overline{R_{avai}}= 100-R_{avai},
\end{align}

where $N_{rep}$ is the number of periodic repetitions (i.e., alternation between normal and attack states). $R_{avai}$ refers to the percentage of time during which the gNB has available resources and is able to respond to incoming requests.


\section{Attack Implementation and Validation}\label{attval} 

\subsection{Attack Implementation with OAI}


To execute the RRC signaling storm attack (pseudo-code in Algorithm~\ref{alg1}), we modified the OAI-UE code to send a Msg1 to the gNB in every frame (note that in the current implementation of the OAI-gNB, there is a limit in the maximum number of Msg1 that can be received per frame). 

The attack is then carried out by continuously looping through Msg1, Msg2, and Msg3, repeatedly initiating the RRC connection procedure. The RACH procedure must be performed before the RRC process and before sending Msg3, as the UE cannot transmit Msg3 without receiving an uplink grant, which is provided by the gNB in Msg2. This ensures that the necessary Random Access (RA) steps are completed before each transmission of Msg3, maintaining the attack loop.

Upon sending a Msg3 to a gNB, a BUE starts a T300 timer, which can range from 100ms to 2000ms, to retransmit Msg3 if Msg4 is not received before T300 expires. In an RRC signaling storm, we assume that the attacker ignores this timer and continuously sends Msg3s at intervals much shorter than the T300 value.

\begin{algorithm}
\caption{RRC Signaling Storm Implementation Steps}
\label{alg1}
\begin{algorithmic}[1]
 \STATE \textbf{Step 1: Modify RRC Behavior}
 \REPEAT
 \STATE Repeatedly flush the Msg3 buffer
 \STATE Prepare Msg3 again
 \STATE Always ignore reception of Msg4
 \UNTIL{Attack stops}

 \STATE \textbf{Step 2: Ignore T300 Timer Expiration}

 \STATE \textbf{Step 3: Trigger RA Procedures on Msg3 Send}
 \IF{Msg3 is sent}
 \STATE Trigger a new RACH
 \STATE Set the flag for Msg2 as not received
 \STATE Reset random access state to generate a preamble
 \ENDIF
\end{algorithmic}
\end{algorithm}

\subsection{Attack Validation}

An attack rate of 132.07 Msg3s/sec was achieved, corresponding to one message every frame (i.e., duration of frame is 7ms). 

To validate the theoretical model presented in Section~\ref{theosec}, four RRC signaling storm attacks were conducted under varying load conditions, i.e., with different numbers of UEs connected to the gNB. Table~\ref{theotable} compares the theoretical and experimental values across four scenarios: no UEs connected ($16$ resources are available for the attacker), $25\%$ resource occupancy ($12$ resources are available for the attacker), $50\%$ resource occupancy ($8$ resources are available for the attacker), and $75\%$ resource occupancy ($4$ resources are available for the attacker). For each scenario, the table lists the number of accepted Msg3s, the number of rejected Msg3s, the drop time, the duration of accept, the duration of reject, and the availability rate. The close alignment between the theoretical and experimental values confirms the accuracy of the model. As expected, the severity of the attack increases (i.e., availability rate decreases) with the number of connected UEs. 
Using a total of $16$ resources at the gNB and a waiting time of $2.7$ seconds, the attacker can render the gNB unavailable for around $\approx 95\%$ of the time. If the gNB is already serving UEs, its capacity would be further strained, leading to an extended period of unavailability.
It is important to note that in certain experimental scenarios (specifically at $50\%$ and $75\%$ loads), the OAI-gNB crashes (i.e., due to the severity of the attack on OAI-gNB), making it impossible to retrieve experimental values (i.e., N/A).


{
\begin{table*}[!htbp]
\vspace*{0.2cm}
\caption{Comparison of theoretical and experimental values for different attack scenarios}
\centering 
\begin{tabular}{|l|c|c|c|c|c|c|c|}
\hline
{\bf Connected UEs} & {\bf Value type} &{\bf $\#$ of accepted}&{\bf $\#$ of rejected} &{\bf Drop time }&{\bf Duration of}&{\bf Duration of}&{\bf Avai. rate} \\
{\bf } & {\bf} &{\bf Msg3s}&{\bf Msg3s} &{\bf }&{\bf accept}&{\bf reject}&{\bf} \\ \hline \hline

\multirow{2}{*}{ $0\%$ ($0/16$)} 	& {Exp. values}	& {$16$}	& {$382$}&{$0.146$s}&{$0.146$s}&{$2.611$s}&{$4.02\%$} \\ \cline{2-8}
& {Theo. values}	& {$16$}	& {$346$}&{$0.121$s}&{$0.121$s}&{$2.636$s}&{$4.42\%$} \\ \hline

\multirow{2}{*}{ $25\%$ ($4/16$)} 	& {Exp. values}	& {$12$}	& {$346$}&{$0.121$s}&{$0.121$s}&{$2.636$s}&{$3.35\%$} \\ \cline{2-8}
& {Theo. values}	& {$12$}	& {$352$}&{$0.091$s}&{$0.091$s}&{$2.666$s}&{$3.29\%$} \\ \hline

\multirow{2}{*}{ $50\%$ ($8/16$)} 	& {Exp. values}	& {$8$}	& {N/A}&{$0.083$s}&{$0.083$s}&{N/A}&{N/A} \\ \cline{2-8}
& {Theo. values}	& {$8$}	& {$356$}&{$0.061$s}&{$0.061$s}&{$2.696$s}&{$2.2\%$} \\ \hline

\multirow{2}{*}{ $75\%$ ($12/16$)} 	& {Exp. values}	& {$4$}	& {N/A}&{$0.021$s}&{$0.021$s}&{N/A}&{N/A} \\ \cline{2-8}
& {Theo. values}	& {$4$}	& {$359$}&{$0.03$s}&{$0.03$s}&{$2.727$s}&{$1.1\%$} \\ \hline

\end{tabular}
\label{theotable}
\end{table*}
}


\section{Proposed Detection Technique}\label{sol} 



To effectively detect RRC signaling storms and distinguish them from high-load scenarios, the initial connection process between a UE and a gNB is analyzed to identify, understand, and detect attack patterns. Since RRC signaling storms occur at the RRC layer, it is essential to consider the relevant features at this layer, as well as those in the lower layers. These features have been examined to determine the most significant and useful ones for the detection system.



To this end, three key features have been identified: the number of Msg3s, $R1$, and $R2$. The first feature, the number of Msg3s, is crucial as it indicates/flags an abnormal event, whether it is an attack or a high-load scenario. In both cases, the gNB would experience a significant increase in the number of Msg3s. Additionally, the gNB will also observe a decrease in the number of Msg5s. During an attack, the Malicious UE (MUE) will not respond to the received Msg4s with a Msg5. In a high-load scenario, the gNB will receive a large number of Msg3s, leading to an overload that prevents it from accommodating all new UEs. Hence, not all UEs will receive a Msg4. The subset of UEs that do receive a Msg4 will send a Msg5, resulting in a reduction in the number of Msg5s.

The second feature, $R1$, represents the ratio of the number of Msg5s to the number of Msg3s (i.e., $R1$ values range between $0$ and $1$). It reflects the proportion of completed (or incomplete) RRC setup procedures. Since the number of Msg5s decreases in both attack and high-load scenarios, $R1$ will also decrease. As a result, this ratio serves as a supportive feature that helps confirm the detection of an abnormal event from the gNB's perspective. However, $R1$ alone is not sufficient to distinguish between an attack and a high-load scenario.

The third feature, $R2$, is defined as the ratio of the number of Msg5s to the number of Msg4s (with $R2$ values ranging between $0$ and $1$). This ratio reflects the proportion of Msg5s received by the gNB in response to the Msg4s it sends. In a high-load scenario, UEs that receive a Msg4 will typically respond with a Msg5, making $R2$ (ideally) equal to $1$. Conversely, during an attack, the MUE will not respond with a Msg5 even after receiving a Msg4, causing $R2$ to decrease and tend toward $0$. Therefore, $R2$ can be used to differentiate between an attack and a high-load scenario. The threshold for both $R1$ and $R2$ is set to $0.5$.


In this research, we use fixed threshold values as an initial step in a simple detection system. While this approach is not adaptive to complex environments, it helps validate the theory and assess the model's feasibility. Future work on adaptive thresholds can build upon the findings of this study.

It is important to note that, alongside the continuous monitoring of the discussed features, capturing the timestamps of the received messages is crucial. This ensures the proper and accurate tracking of feature variations over time.



The proposed features will exhibit distinct values and varying trends depending on the state of the gNB, whether it is in a normal, attack, or high-load condition. These differences can be leveraged to develop an RRC signaling storm detection system. The trend (or profile) associated with each state is detailed in Table~\ref{trends}.

{
\begin{table}[!htbp]
\caption{Feature trends for different states}
\centering 
\begin{tabular}{|l|c|c|c|}
\hline
{\bf State} & {\bf $\#$ of Msg3s} &{\bf $R1$}&{\bf $R2$} \\ \hline \hline
{ Attack} 	& {$\uparrow$}	& {$\downarrow$}	& {$\downarrow$} \\ \hline
{ High-Load} 	& {$\uparrow$}& {$\downarrow$}	& {$=$ (i.e., 1)} \\ \hline
\end{tabular}
\label{trends}
\end{table}
}






\section{Detection Solution Implementation and Evaluation in O-RAN}\label{deteval}



\subsection{Environment}\label{env}




The simulated system runs on a virtual machine with Ubuntu 20.04.6 and 12 GB of RAM.

In the implemented scenario, a single OAI MUE targets the gNB, generating an attack rate of $132$ Msg3s per second with a waiting time of $2.7$ seconds. To simulate normal conditions with background traffic from BUEs directed toward the base station, a Truncated Poisson Distribution is employed. This distribution periodically adds a new number of RRC messages to the counters in the xApp, mimicking normal traffic behavior. The distribution has a mean of $2$, with lower and upper bounds of $0$ and $3$, respectively. The OAI-gNB supports up to $16$ UEs (i.e., $16$ resources) simultaneously. A sliding window of $625$ms is used for detection. 

\begin{figure*}[!ht]
\centering
\subfloat[][]{\includegraphics[width=\textwidth]{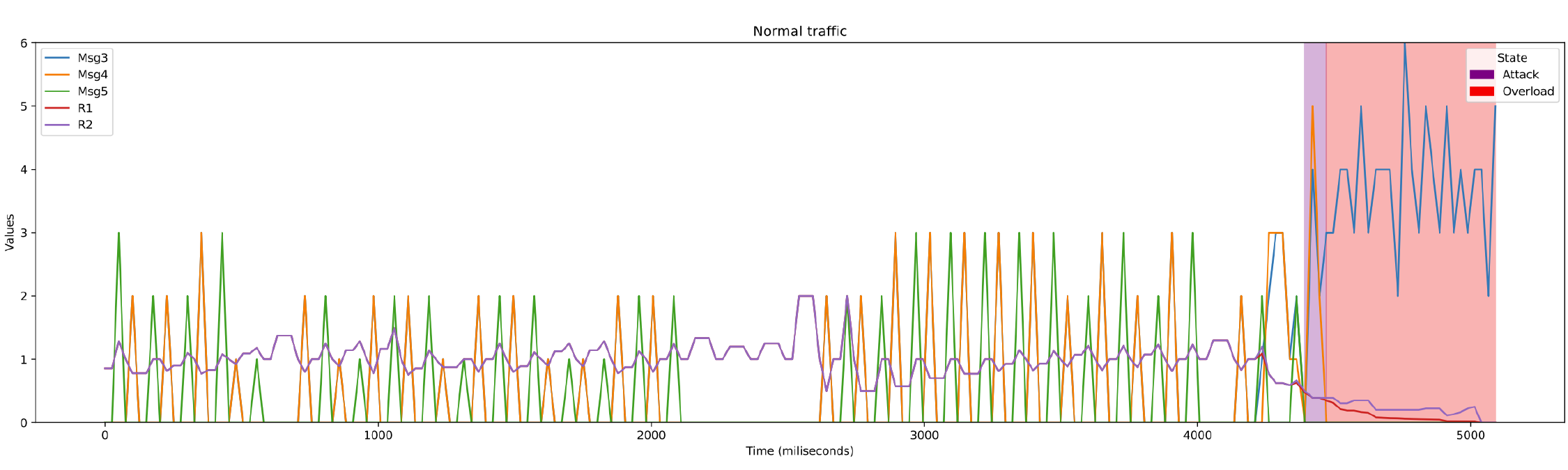}\label{norm1}}\\
\subfloat[][]{\includegraphics[width=0.49\textwidth]{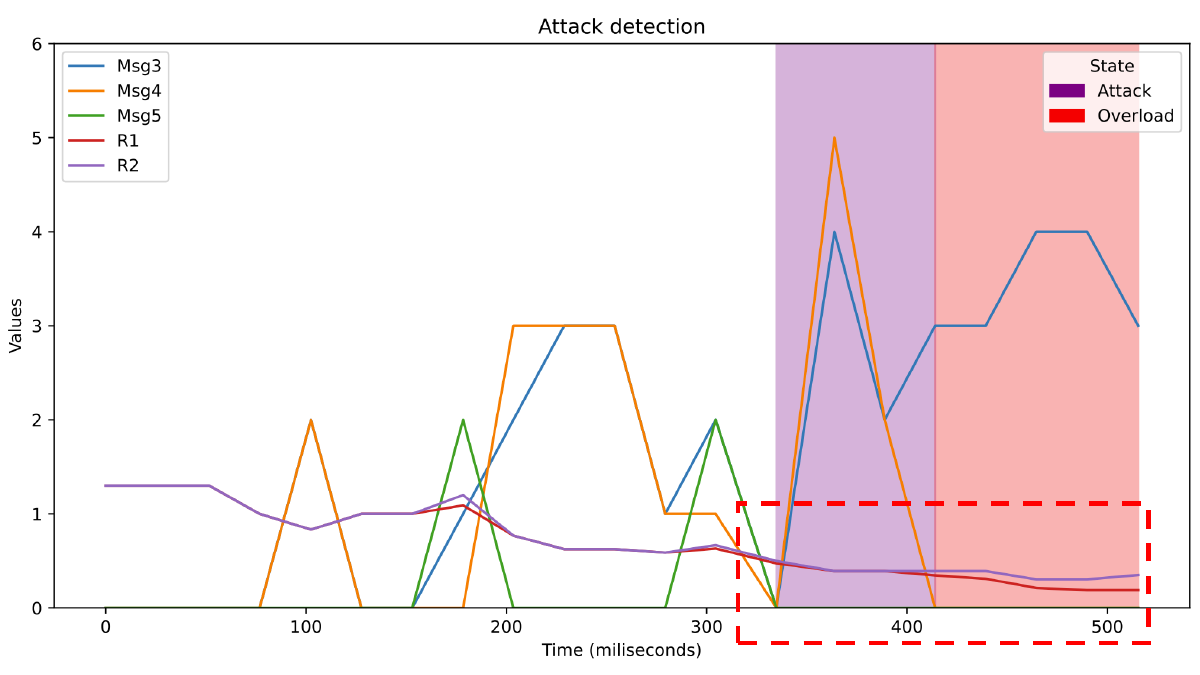}\label{norm2}}\hfill
\subfloat[][]{\includegraphics[width=0.49\textwidth]{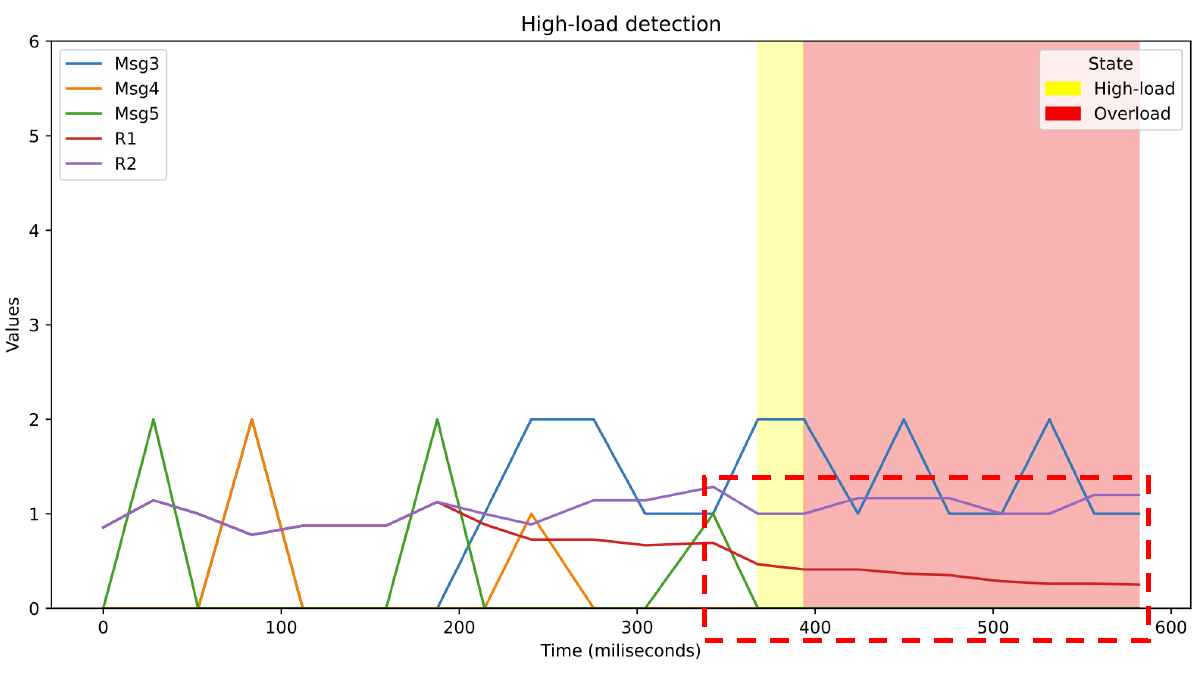}\label{hl}}
\caption{Performance of the proposed detection solution under (a) normal traffic conditions, (b) attack, and (c) high-load}
\end{figure*}

\subsection{Results and Discussion} \label{results}

The evaluation focuses on two key aspects, performance (impact) of the proposed detection solution and its effectiveness, under two test scenarios, attack and high-load.

\textbf{-- Impact of Detection Solution:} To assess the performance of the proposed detection solution, we measured its impact on resource usage and the associated overhead. Key performance metrics include CPU (Central Processing Unit) and memory consumption. CPU consumption refers to the amount of processing power utilized by the detection solution, while memory consumption indicates the amount of used RAM (Random Access Memory). 
The detection system, which comprises FlexRIC and the xApp, demonstrates minimal resource usage compared to the overall OAI system. It utilizes approximately 1.2\% of the CPU and nearly 0\% of the memory, showcasing a highly efficient performance with minimal impact on processing and computing power. Similar results were observed for the high-load scenario.



\textbf{-- Detection Solution Effectiveness:} 
The effectiveness of the solution, particularly during an attack, has been evaluated by measuring detection time, or latency. Detection time refers to how quickly the solution can identify abnormal activity and differentiate it from a high-load situation, signaling an attack. Ideally, this latency should be minimized to ensure swift detection. 

Figure~\ref{norm1} illustrates the incoming messages to a gNB over time, incorporating values from the Truncated Poisson distribution. An attack is initiated towards the end of the monitored period. Messages Msg3, Msg4, and Msg5 are added, periodically. The plots of Msg3 and Msg4 overlap due to their simultaneous occurrence, while Msg5 exhibits a delay due to the natural lag between Msg4 and Msg5.
The values of R1 and R2 are determined based on the number of messages within a sliding window. During normal traffic simulation, the R1 and R2 ratios fluctuate around 1 without dropping significantly below the thresholds, thus avoiding any alerts. However, towards the end of the period, a substantial increase in incoming Msg3s occurs, generated by the MUE with the highest possible attack rate. Figure~\ref{norm2} provides a detailed view of the attack phase. 
The detection state has four possible values representing the gNB’s status: Normal, Attack, High-load, and Overload. The figure precisely depicts the behavior of incoming RRC messages. Initially, the gNB responds with some Msg4s to the attacker’s Msg3s but does not receive any Msg5s. Eventually, as resources are exhausted, the gNB stops sending Msg4s, though it continues receiving a high volume of Msg3s. Throughout the attack, R1 and R2 values progressively decline, eventually reaching 0, prompting the gNB’s detected state to shift from Normal to Attack, and then to Overload. This transition occurs shortly after the attack begins. 


\begin{figure}[!ht]
\centering
\includegraphics[width=0.5\textwidth]{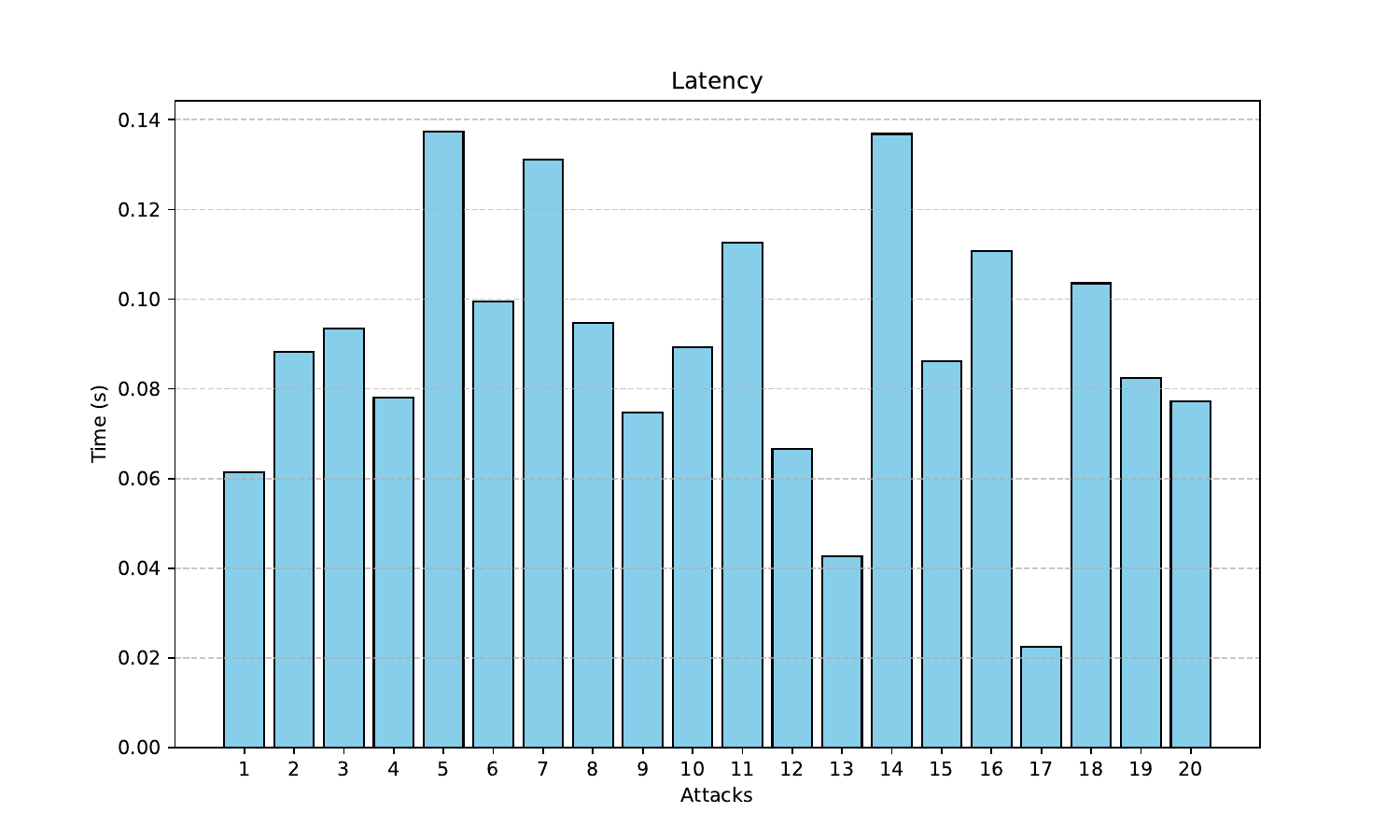}
\caption{Detection latency (i.e., attack case)}
\label{latency}
\end{figure}

To evaluate detection latency, the RRC signalling storm has been executed 20 times with an attack rate of $132$ Msg/sec. On average, attacks are detected in about $\approx 90$ms, which is $\approx 60$ms before the gNB becomes overloaded (i.e., drop time is equal to $\approx 150$ms, thus, $150-90 \approx 60$ms) (Figure~\ref{latency}). This detection latency is considered acceptable since it enables the system to identify attack attempts before the gNB reaches overload or becomes unavailable. This early detection provides a short window for implementing mitigation measures to safeguard the gNB's availability.


On the other hand, Figure~\ref{hl} shows the variation of features over time for the high-load test case. The proposed detection solution accurately identifies when the gNB is experiencing a high load and when it becomes unavailable or blocked (i.e., enters the overload state). Throughout the simulated scenario, the value of R1 progressively decreases until it eventually reaches 0, while the value of R2 remains close to 1. The proposed detection system can effectively distinguish between an attack and a high-load. On average, high-load scenarios are detected within $131$ms.

\section{Conclusion}\label{conc}

In this paper, we implemented an RRC signaling storm and assessed its impact using OpenAirInterface (OAI). Our experiments revealed an attacker rate of approximately 132 Msg3s/sec, rendering the gNB unavailable 95\% of the time. The observed impact aligns closely with our proposed theoretical model, confirming its accuracy. 
To address the challenge of RRC signaling storms within the O-RAN architecture, we introduced a threshold-based detection solution that leverages RRC layer characteristics to effectively distinguish between benign high-load scenarios and malicious attacks. Our validation results demonstrate that this solution can reliably detect (i.e., within $\approx 90$ms) and potentially mitigate RRC signaling storms with minimal resource overhead (i.e., negligible CPU and memory consumption), significantly enhancing network resilience and reliability. This approach strengthens 5G network security by providing robust protection against disruptive signaling storms while maintaining low resource impact. 


As next steps, we will focus on refining and enhancing the detection mechanism, expanding our evaluation to encompass a broader range of scenarios, and developing effective mitigation techniques. 
Additionally, we aim to explore methods for predicting optimal detection thresholds to better anticipate and prevent future signaling storms.
An adaptive threshold-based algorithm can be studied and used to improve the detection performance and increase the flexibility of the detection technique in different systems and environment.

\bibliographystyle{IEEEtran}
\bibliography{ref}

\vspace{12pt}

\end{document}